\documentclass[conference]{IEEEtran}
\usepackage{amssymb}
\usepackage[cmex10]{amsmath}
\usepackage{stfloats}
\usepackage{graphicx}
\usepackage{subfigure}
\usepackage{tabularx}
\usepackage{epsfig,epsf,color,balance,cite}
\usepackage{verbatim}
\usepackage{url}
\usepackage{bm}
\usepackage{mathrsfs}
\usepackage{amsthm}
\DeclareMathAlphabet\mathbfcal{OMS}{cmsy}{b}{n}

\newtheorem{theorem}{Theorem}

\newtheorem{lemma}{Lemma}
\newtheorem{remark}{Remark}
\usepackage{algorithm}
\usepackage{algpseudocode}
\usepackage{graphicx}
\usepackage{epstopdf}
\IEEEoverridecommandlockouts
	
\begin{document}

\title{SD-Based Low-Complexity Precoder Design for Gaussian MIMO Wiretap Channels}

\author{
	\IEEEauthorblockN{Hao Xu\IEEEauthorrefmark{1},    
		Kai-Kit Wong\IEEEauthorrefmark{1},
		and
		Giuseppe Caire\IEEEauthorrefmark{2}
	}
	\IEEEauthorblockA{\IEEEauthorrefmark{1}Department of Electronic and Electrical Engineering, University College London, London WC1E7JE, U.K.}
	\IEEEauthorblockA{\IEEEauthorrefmark{2}Faculty of Electrical Engineering and Computer Science, Technical University of Berlin, 10587 Berlin, Germany}
	\IEEEauthorblockA{E-mail: hao.xu@ucl.ac.uk; kai-kit.wong@ucl.ac.uk; caire@tu-berlin.de}
}

\maketitle

\begin{abstract}
This paper considers a Gaussian multi-input multi-output (MIMO) multiple access wiretap (MAC-WT) channel, where an eavesdropper (Eve) wants to extract the confidential information of all users.
Assuming that both the legitimate receiver and Eve jointly decode their interested messages, we aim to maximize the sum secrecy rate of the system by precoder design.
Although this problem could be solved by first using the iterative majorization minimization (MM) based algorithm to get a sequence of convex log-determinant optimization subproblems and then using some general tools, e.g., the interior point method, to deal with each subproblem, this strategy involves quite high computational complexity.
Therefore, we propose a simultaneous diagonalization based low-complexity (SDLC) method to maximize the secrecy rate of a simple one-user wiretap channel, and then use this method to iteratively optimize the covariance matrix of each user.
Simulation results show that in contrast to the existing approaches, the SDLC scheme achieves similar secrecy performance but requires much lower complexity.
\end{abstract}

\IEEEpeerreviewmaketitle

\section{Introduction}
\label{section1}

To meet the tremendous demand for wireless communications, the future mobile systems will incorporate many different network topologies and large numbers of devices which may access and leave at any time, making it difficult to generate and manage cryptographic keys. 
In addition, the unprecedented growth of computational ability makes it possible for eavesdroppers (Eves) extracting the confidential information of authorized users without secret keys.
Hence, the conventional cryptographic encryption methods, which rely on secret keys and assumptions of limited computational ability at Eves, are no longer sufficient to guarantee secrecy in the future mobile networks.
Starting from some early seminal works \cite{shannon1949communication, wyner1975wire, csiszar1978broadcast}, the study of information theoretic secrecy in communications has triggered considerable research interests recently \cite{7762075, hao2018resource, 8895802}.

Different from the cryptographic encryption methods employed in the application layer, physical layer security techniques exploit the random propagation properties of radio channels and advanced signal processing techniques to prevent Eves from wiretapping.
Over the past decades, the multiple access wiretap (MAC-WT) channels have drawn great research interests \cite{ekrem2008secrecy, tekin2008gaussian, nafea2019generalizing, 9174164, tekin2008general, xu2022achievable}.
Although the secrecy capacity regions of MAC-WT channels are still unknown, the outbounds and achievable regions have been widely studied for different MAC-WT cases, e.g., the case with a weaker Eve which has access to a degraded version of the main channel \cite{ekrem2008secrecy, tekin2008gaussian}, the non-degraded case with different wiretapping scenarios \cite{nafea2019generalizing}, the non-degraded case where each user has both confidential and opens message intended for the legitimate receiver \cite{9174164, tekin2008general, xu2022achievable}, etc.

Based on the information theoretic results, a lot of work further studied the resource allocation problems in MAC-WT channels \cite{tekin2008general, lee2017precoder, xu2022achievable}.
The sum secrecy rate of a Gaussian single-input single-output (SISO) MAC-WT channel was maximized by power control in \cite{tekin2008general}.
Reference \cite{lee2017precoder} maximized the sum secrecy rate of a Gaussian multi-input multi-output (MIMO) MAC-WT system, but considered a special power constraint, making the secrecy performance limited.
The same problem as in \cite{lee2017precoder} but with a general power constraint was considered in \cite{xu2022achievable} (see \cite[Problem~$(23)$]{xu2022achievable}) and the iterative majorization minimization (MM) based scheme was applied to solve this problem, which is a difference of convex (DC) programming.
As shown in \cite[Fig.~$6$]{xu2022achievable}, the system secrecy performance can be greatly improved in contrast to \cite{lee2017precoder}.
However, as analyzed in \cite[Subsection~IV-C]{xu2022achievable}, the MM-based scheme involves quite high computational complexity and it becomes prohibitive to perform this scheme when the network size is large.

In this paper, we consider a Gaussian MIMO MAC-WT system and aim to maximize the sum secrecy rate, i.e., again solve \cite[Problem~$(23)$]{xu2022achievable}. 
Notice that Gaussian coding with specific spatial covariance matrix (in the antenna dimension) can be obtained by `coloring' an independent and identically distributed (i.i.d.) Gaussian signal by linear spatial precoding. 
This problem is also referred to as {\em precoder design}.
To reduce the computational complexity in solving this problem, motivated by the iterative water-filling method \cite[Subsection~$9.2$]{el2011network}, we iteratively optimize the signal covariance matrix of each user.
It is shown that when all the other users' covariance matrices are fixed, the original problem can be equivalently transformed to the secrecy rate maximization problem of a simple one-user wiretap channel.
Hence, we first consider a single-user MIMO wiretap channel and propose a simultaneous diagonalization based low-complexity (SDLC) scheme, and then solve the original problem by iteratively applying this scheme.
Note that as a general method, besides the sum secrecy rate maximization problem considered in this paper, the SDLC scheme proposed here can be applied to deal with a variety of problems whose intermediate steps can be formulated as the maximization of a difference of log-determinants.

We have to point out that the secrecy rate maximization problem of a one-user wiretap channel has been widely studied and the analytical capacity-achieving solution exists for some special cases, e.g., single-transmit-antenna case \cite{parada2005secrecy}, single-receive-antenna case \cite{khisti2010secureI}, two-transmit-antenna case \cite{shafiee2009towards, vaezi2017optimal}, high SNR case \cite{khisti2010secure}, etc.
However, the analytical solution for the general MIMO wiretap channel is still an open problem.
In \cite{fakoorian2012optimal}, the generalized singular value decomposition (GSVD) was applied to decompose the MIMO wiretap channel into a set of parallel sub-channels and a sub-optimal solution was obtained.
In contrast to \cite{fakoorian2012optimal}, we provide more insightful analysis, show that the proposed SDLC scheme is determined by the channel state, and give the uniqueness condition.
If this condition is satisfied, the SDLC scheme is unique and is equivalent to the GSVD scheme in terms of the secrecy rate.
Otherwise, we can get many different SDLC schemes.
Moreover, we show by simulation that compared with the MM-based scheme provided in \cite{xu2022achievable} and the GSVD method given in \cite{fakoorian2012optimal}, the proposed SDLC method achieves similar secrecy performance but involves much lower computational complexity.


\section{System Model and Problem Formulation}
\label{GV_MAC-WT}

Consider a Gaussian MIMO MAC-WT channel with $K$ users, a legitimate receiver (or Bob for brevity), and an Eve.
Each user $k$, Bob, and Eve are respectively equipped with $T_k$, $B$, and $E$ antennas.
Let ${\bm x}_k \in {\mathbb C}^{T_k \times 1}$ denote the signal vector of user $k$ and assume Gaussian channel input, i.e., $\bm x_k \sim {\cal CN}(\bm 0, \bm F_k)$, where the covariance matrix $\bm F_k$ has power constraint ${\text {tr}}(\bm F_k) \leq P_k$.
The received signals at Bob and Eve are given by
\begin{align}\label{GV_YZ}
& \bm y = \sum_{k=1}^K \bm H_k \bm x_k + \bm n_{\text B},\nonumber\\
& \bm z = \sum_{k=1}^K \bm G_k \bm x_k + \bm n_{\text E},
\end{align}
where $\bm H_k \in {\mathbb C}^{B \times T_k}$ and $\bm G_k \in {\mathbb C}^{E \times T_k}$ are constant channel gain matrices from user $k$ to Bob and Eve, and $\bm n_{\text B} \in {\mathbb C}^{B \times 1}$ and $\bm n_{\text E} \in {\mathbb C}^{B \times 1}$ are additive Gaussian noise vectors at Bob and Eve with $\bm n_{\text B} \sim {\cal CN}(0, \sigma_B^2 \bm I_B)$ and $\bm n_{\text E} \sim {\cal CN}(0, \sigma_E^2 \bm I_E)$.
Assume that both Bob and Eve jointly decode their interested messages.
The achievable regions for such a MIMO MAC-WT channel has be studied in \cite{xu2022achievable} and the maximum achievable sum secrecy rate of the system is \cite[$(20)$]{xu2022achievable}
\begin{align}\label{R_s_joint_GV}
& R (\bm F_{\cal K}) = \left[ I(\bm x_{\cal K}; \bm y) - I(\bm x_{\cal K}; \bm z) \right]^+ \nonumber\\
& =\!\! \left[\!\log\! \left| \sum_{k=1}^K\! \frac{1}{\sigma_B^2} \bm H_k \bm F_k \bm H_k^H \!\!+\! \bm I_B \right| \!-\! \log\! \left| \sum_{k=1}^K\! \frac{1}{\sigma_E^2} \bm G_k \bm F_k \bm G_k^H \!\!+\! \bm I_E \right|\right]^{\!\!+}\!\!\!.
\end{align}
Note that in this paper we use calligraphic subscript to denote the set of elements whose indexes take values from the subscript set, e.g., $\bm F_{\cal K} = \left\{ \bm F_1, \cdots, \bm F_K \right\}$ and $\bm x_{\cal K} = \{\bm x_k, \forall k \in {\cal K}\}$ in (\ref{R_s_joint_GV}).
We aim to maximize $R (\bm F_{\cal K})$ by designing the covariance matrices.
The problem can be formulated as
\begin{subequations}\label{C1_max}
	\begin{align}
	\mathop {\max }\limits_{\bm F_{\cal K}} \quad & R (\bm F_{\cal K}) \label{C1_max_a}\\
	\text{s.t.} \quad\; &  {\text {tr}}(\bm F_k) \leq P_k, ~\forall~ k \in \cal K, \label{C1_max_b}\\
	& \bm F_k \succeq \bm 0, ~\forall~ k \in \cal K. \label{C1_max_c}
	\end{align}
\end{subequations}

We have studied problem (\ref{C1_max}) in \cite{xu2022achievable}.
Since it is a DC programming, we obtain a sub-optimal solution in \cite{xu2022achievable} by using \cite[Algorithm~$1$]{xu2022achievable}, which is an iterative MM-based algorithm and solves a sequence of convex log-determinant optimization subproblems using some general tools, e.g., interior point method, the CVX tools provided by Matlab, etc.
However, as analyzed in \cite[Subsection~IV-C]{xu2022achievable} and shown by the simulation results in this paper, \cite[Algorithm~$1$]{xu2022achievable} involves quite a high complexity.
It is very time-consuming to execute \cite[Algorithm~$1$]{xu2022achievable} and will become even impractical when the network size grows large.
Hence, we aim to find an efficient and low-complexity method to solve (\ref{C1_max}).
Motivated by the iterative water-filling method \cite[Subsection~$9.2$]{el2011network}, we iteratively optimize the covariance matrices of all users, i.e., $\bm F_k, \forall k \in {\cal K}$, and hope that we could solve the corresponding problem in each step with a low complexity.

\section{Single-user Gaussian MIMO Wiretap Channel}
\label{one_UE}

Before solving (\ref{C1_max}), we first consider a single-user Gaussian MIMO wiretap channel and give the SDLC scheme for this simple case.
With one transmitter, problem (\ref{C1_max}) reduces to
\begin{subequations}\label{problem_one_UE}
	\begin{align}
	\mathop {\max }\limits_{\bm F} \, & \log\! \left| \bm H \bm F \bm H^H \bm \varOmega_1^{-1} \!+\! \bm I_B \right| \!-\! \log\! \left| \bm G \bm F \bm G^H \bm \varOmega_2^{-1} \!+\! \bm I_E \right| \label{problem_one_UE_a}\\
	\text{s.t.} \,\, &  \bm F \succeq \bm 0, \label{problem_one_UE_b}\\
	& {\text {tr}}(\bm F) \leq P, \label{problem_one_UE_c}
	\end{align}
\end{subequations}
where we omit the user index and $\left[ \cdot \right]^+$ in (\ref{problem_one_UE_a}) for brevity.
Note that to solve (\ref{C1_max}) using the SDLC scheme proposed in this section, we replace $\sigma_B^2 \bm I_B$ and $\sigma_E^2 \bm I_E$ with the general noise covariance matrices $\bm \varOmega_1$ and $\bm \varOmega_2$, which are assumed to be positive definite.
As explained in the introduction part, problem (\ref{problem_one_UE}) has been widely studied but its analytical solution is still an open problem.
In the following we provide a SDLC scheme, with which the MIMO wiretap channel can be decomposed into a set of parallel sub-channels and the confidential information can be transmitted over sub-channels where Bob experiences better channel state than Eve.
In contrast to the GSVD scheme given in \cite{fakoorian2012optimal}, which also decomposes the channel, we show that the SDLC scheme is determined by the channel state and give the uniqueness condition.
If this condition is satisfied, the SDLC scheme is unique and is equivalent to the GSVD scheme in terms of the secrecy rate.
Though, in this case, equivalent in secrecy performance, we show by simulation that compared with the GSVD scheme, the computational complexity can be greatly decreased by the SDLC scheme.
If the uniqueness condition is not satisfied, we can get many different SDLC schemes and each one may provide a different solution to (\ref{problem_one_UE}).

Before introducing the SDLC scheme, we first simplify the objective function (\ref{problem_one_UE_a}).
Denote the eigendecomposition of $\bm \varOmega_1$ and $\bm \varOmega_2$ by $\bm \varOmega_1 = \bm \varGamma_1 \bm \varLambda_1 \bm \varGamma_1^H$ and $\bm \varOmega_2 = \bm \varGamma_2 \bm \varLambda_2 \bm \varGamma_2^H$, respectively.
Then, using the fact that $\left| \bm O_1 \bm O_2 + \bm I \right| = \left| \bm O_2 \bm O_1 + \bm I \right|$, the first term of (\ref{problem_one_UE_a}) can be rewritten as
\begin{align}\label{rewrite_term1}
\log \!\left| \bm H \bm F \bm H^H \bm \varOmega_1^{-1} \!+\! \bm I_B \right| \!=&  \log \!\left| \bm \varLambda_1^{-\frac{1}{2}} \!\bm \varGamma_1^H \bm H \bm F \bm H^H \bm \varGamma_1 \bm \varLambda_1^{-\frac{1}{2}} \!+\! \bm I_B \right| \nonumber\\
= & \frac{1}{\ln 2} \ln \left| {\bm {\hat H}} \bm F {\bm {\hat H}}^H + \bm I_B \right|,
\end{align}
where ${\bm {\hat H}} = \bm \varLambda_1^{-\frac{1}{2}} \bm \varGamma_1^H \bm H$.
Similarly, the second term of (\ref{problem_one_UE_a}) can be rewritten as
\begin{equation}\label{rewrite_term2}
\log \left| \bm G \bm F \bm G^H \bm \varOmega_2^{-1} + \bm I_E \right| = \frac{1}{\ln 2} \ln \left| {\bm {\hat G}} \bm F {\bm {\hat G}}^H + \bm I_E \right|,
\end{equation}
where ${\bm {\hat G}} = \bm \varLambda_2^{-\frac{1}{2}} \bm \varGamma_2^H \bm G$.
Problem (\ref{problem_one_UE}) can then be equivalently transformed to
\begin{subequations}\label{general_problem_1}
	\begin{align}
	\mathop {\min }\limits_{\bm F} \quad & - \ln \left| {\bm {\hat H}} \bm F {\bm {\hat H}}^H + \bm I_B \right| + \ln \left| {\bm {\hat G}} \bm F {\bm {\hat G}}^H + \bm I_E \right| \label{general_problem_1_a}\\
	\text{s.t.} \quad\, &  ({\text{\ref{problem_one_UE_b}}}),~ ({\text{\ref{problem_one_UE_c}}}).
	\end{align}
\end{subequations}
We focus on solving problem (\ref{general_problem_1}) in the following.

Since ${\bm {\hat H}}^H {\bm {\hat H}}$ and ${\bm {\hat G}}^H {\bm {\hat G}}$ are both positive semi-definite matrices, for any vector $\bm d \in {\mathbb C}^{T \times 1}$, if $\bm d^H ({\bm {\hat H}}^H {\bm {\hat H}} + {\bm {\hat G}}^H {\bm {\hat G}}) \bm d = 0$, there must be $\bm d^H {\bm {\hat H}}^H {\bm {\hat H}} \bm d = 0$.
It is thus known from \cite[Lemma~2]{au1971note} that ${\bm {\hat H}}^H {\bm {\hat H}} + {\bm {\hat G}}^H {\bm {\hat G}}$ and ${\bm {\hat H}}^H {\bm {\hat H}}$ can be simultaneously diagonalized.
In particular, there exists a non-singular matrix $\bm U_1$ such that
\begin{align}\label{UHHGGU}
\bm U_1^H ({\bm {\hat H}}^H {\bm {\hat H}} + {\bm {\hat G}}^H {\bm {\hat G}}) \bm U_1 = \begin{bmatrix}
\bm I_{T_0} & \bm 0\\
\bm 0 & \bm 0
\end{bmatrix},\nonumber\\
\bm U_1^H {\bm {\hat H}}^H {\bm {\hat H}} \bm U_1 = 
\begin{bmatrix}
\bm W & \bm 0\\
\bm 0 & \bm 0
\end{bmatrix},
\end{align}
where $T_0 = {\text {rank}} ({\bm {\hat H}}^H {\bm {\hat H}} + {\bm {\hat G}}^H {\bm {\hat G}})$ and $\bm W \in {\mathbb C}^{T_0 \times T_0}$.
To construct $\bm U_1$, denote the eigendecomposition of ${\bm {\hat H}}^H {\bm {\hat H}} + {\bm {\hat G}}^H {\bm {\hat G}}$ by 
\begin{equation}\label{eigende_HHGG}
{\bm {\hat H}}^H {\bm {\hat H}} + {\bm {\hat G}}^H {\bm {\hat G}} = \bm \varPsi_1 \begin{bmatrix}
\bm \varUpsilon & \bm 0\\
\bm 0 & \bm 0
\end{bmatrix} \bm \varPsi_1^H,
\end{equation}
where $\bm \varPsi_1 \in {\mathbb C}^{T \times T}$ is a unitary matrix and $\bm \varUpsilon \in {\mathbb R}^{T_0 \times T_0}$ is a diagonal matrix with positive diagonal entries.
Let
\begin{equation}\label{U1}
\bm U_1 = \bm \varPsi_1 
\begin{bmatrix} \bm \varUpsilon^{- \frac{1}{2}} & \bm 0\\ \bm 0 & \bm \varPi_1 \end{bmatrix},
\end{equation}
where $\bm \varPi_1$ can be any square matrix of dimension $T - T_0$.
It is obvious that the $\bm U_1$ resulted from (\ref{U1}) guarantees (\ref{UHHGGU}).
Since $\bm U_1^H {\bm {\hat H}}^H {\bm {\hat H}} \bm U_1 \succeq \bm 0$ and
\begin{equation}\label{UGGU}
\bm U_1^H {\bm {\hat G}}^H {\bm {\hat G}} \bm U_1 = 
\begin{bmatrix}
\bm I_{T_0} - \bm W & \bm 0\\
\bm 0 & \bm 0
\end{bmatrix} \succeq \bm 0,
\end{equation}
it is known that
\begin{equation}\label{IG0}
\bm I_{T_0} \succeq \bm W \succeq \bm 0.
\end{equation}
Denote the eigendecomposition of $\bm W$ by
\begin{equation}\label{decomp_G0}
\bm W = \bm \varPsi_2 {\text {diag}} \{ \rho_1, \cdots, \rho_{T_0} \} \bm \varPsi_2^H,
\end{equation}
where $\bm \varPsi_2 \in {\mathbb C}^{T_0 \times T_0}$ is a unitary matrix and $0 \leq \rho_t \leq 1, ~\forall~ 1 \leq t \leq T_0$.
Let 
\begin{equation}\label{U2}
\bm U_2 = 
\begin{bmatrix}
\bm \varPsi_2 & \bm 0\\
\bm 0 & \bm \varPi_2
\end{bmatrix},
\end{equation}
and 
\begin{equation}\label{U1U2}
\bm U = \bm U_1 \bm U_2,
\end{equation}
where $\bm \varPi_2$ can be any square matrix of dimension $T - T_0$.
${\bm {\hat H}}^H {\bm {\hat H}}$ and ${\bm {\hat G}}^H {\bm {\hat G}}$ can then be simultaneously diagonalized as follows
\begin{align}\label{SDHHGG}
\bm U^H {\bm {\hat H}}^H {\bm {\hat H}} \bm U & = {\text {diag}} \{ \rho_1, \cdots, \rho_{T_0}, 0, \cdots, 0 \}, \nonumber\\
\bm U^H {\bm {\hat G}}^H {\bm {\hat G}} \bm U & = {\text {diag}} \{ 1 - \rho_1, \cdots, 1 - \rho_{T_0}, 0, \cdots, 0 \},
\end{align}
where the last $T - T_0$ diagonal entries of $\bm U^H {\bm {\hat H}}^H {\bm {\hat H}} \bm U$ and $\bm U^H {\bm {\hat G}}^H {\bm {\hat G}} \bm U$ are $0$.
Let 
\begin{equation}\label{F4}
\bm F = \bm U \bm A \bm U^H,
\end{equation}
where $\bm A \triangleq {\text {diag}} \{ a_1, \cdots, a_T \}$ is a diagonal matrix with non-negative real diagonal entries.
The objective function (\ref{general_problem_1_a}) and ${\text {tr}}(\bm F)$ in constraint (\ref{problem_one_UE_c}) can thus be transformed to
\begin{align}\label{SD_ob2}
& - \ln \left| {\bm {\hat H}} \bm F {\bm {\hat H}}^H + \bm I_B \right| + \ln \left| {\bm {\hat G}} \bm F {\bm {\hat G}}^H + \bm I_E \right| \nonumber\\
= & - \ln \left| \bm U^H {\bm {\hat H}}^H {\bm {\hat H}} \bm U \bm A + \bm I_B \right| + \ln \left| \bm U^H {\bm {\hat G}}^H {\bm {\hat G}} \bm U \bm A + \bm I_E \right| \nonumber\\
= & \sum_{t=1}^{T_0} \left[ - \ln (\rho_t a_t + 1) + \ln ((1 - \rho_t) a_t + 1) \right],
\end{align}
and
\begin{align}\label{diag_trace_F2}
{\text {tr}}(\bm F) = {\text {tr}}(\bm U \bm A \bm U^H) = {\text {tr}}( \bm U^H \bm U \bm A) = \sum_{t=1}^T \left\| \bm u_t \right\|^2 a_t,
\end{align}
where $\bm u_t$ is the $t$th column of matrix $\bm U$.
Accordingly, instead of directly solving problem (\ref{general_problem_1}), we consider the following problem and then obtain $\bm F$ from (\ref{F4})
\begin{subequations}\label{SD_problem3}
	\begin{align}
	\mathop {\min }\limits_{\bm A} \quad & \sum_{t=1}^{T_0} \left[ - \ln ( \rho_t a_t + 1) + \ln ((1 - \rho_t) a_t + 1) \right] \label{SD_problem3_a}\\
	\text{s.t.} \quad\; &  a_t \geq 0, ~\forall~ t \in {\cal T}, \label{SD_problem3_b}\\
	& \sum_{t=1}^{T_0} \left\| \bm u_t \right\|^2 a_t \leq P. \label{SD_problem3_c}
	\end{align}
\end{subequations}

By simultaneously diagonalizing ${\bm {\hat H}}^H {\bm {\hat H}}$ and ${\bm {\hat G}}^H {\bm {\hat G}}$ in (\ref{SDHHGG}), the MIMO wiretap channel is decomposed into $T_0$ parallel sub-channels with $\rho_t$ and $1 - \rho_t, \forall t \in {\cal T}_0$, respectively, being the channel gains experienced by Bob and Eve.
$a_t$ can be seen as the power allocated to the $t$th sub-channel.
Since $- \ln (\rho_t a_t + 1)$ and $\ln ((1 - \rho_t) a_t + 1)$ are respectively convex and concave functions of $a_t$, problem (\ref{SD_problem3}) is non-convex.
However, we show in the following theorem and Appendix~\ref{prove_theorem_optimal_a} that $a_t$ is non-zero only when Bob observes a better channel state than Eve, i.e., $\frac{1}{2} < \rho_t \leq 1 $, and the optimal solution can be efficiently obtained.

\begin{theorem}\label{theorem_optimal_a}
	The optimal solution of problem (\ref{SD_problem3}) is
	\begin{equation}\label{optimal_A}
	a_t^* \!=\!\! \left\{\!\!\!\!
	\begin{array}{ll}
	0, ~{\text {if}}~ T_0 \leq t \leq T ~{\text {or}}~ t \in {\cal T}_0 \setminus {\cal J}, \vspace{0.3em} \\
	\left[ \frac{1}{\beta^* \left\| \bm u_t \right\|^2} -1 \right]^+, ~{\text {if}}~ t \!\in\! {\cal J} ~{\text {and}}~ \rho_t = 1, \vspace{0.3em} \\
	\frac{\left[ - 1 + \sqrt{1 - 4 \rho_t (1 - \rho_t) \left( 1 + \frac{1 - 2 \rho_t}{\beta^* \left\| \bm u_t \right\|^2} \right)} \right]^{\!+\!\!}}{2 \rho_t (1 - \rho_t)}, {\text {if}}~\! t \!\in\! {\cal J} \!\!~{\text {and}}~\! \frac{1}{2} \!<\! \rho_t \!<\! 1,
	\end{array} \right.
	\end{equation}
	where ${\cal T}_0 \!=\! \{1, \cdots, T_0\}$, ${\cal J} \!=\! \{ t| 1 \leq t \leq T_0,~ \frac{1}{2} < \rho_t \leq 1 \}$, and $\beta^*$ can be found using the bisection searching method such that the constraint (\ref{SD_problem3_c}) holds with equality.
\end{theorem}
\itshape \textbf{Proof:}  \upshape
See Appendix \ref{prove_theorem_optimal_a}.
\hfill $\Box$

Based on Theorem~\ref{theorem_optimal_a}, a solution of problem (\ref{general_problem_1}) can be obtained by using (\ref{F4}) and (\ref{optimal_A}).

\begin{remark}\label{not_optimal}
	Note that though problem (\ref{SD_problem3}) can be optimally solved, the corresponding solution of (\ref{general_problem_1}) obtained from Theorem~\ref{theorem_optimal_a} and (\ref{F4}) is not necessarily optimal since the formation of $\bm F$ is limited by (\ref{F4}).
\end{remark}

As shown above, for a given channel state, the SDLC scheme is determined by the values of $\rho_t, \forall t \in {\cal T}_0$, which are the eigenvalues of $\bm U_1^H {\bm {\hat H}}^H {\bm {\hat H}} \bm U_1$, and $\bm U$, which is determined by the choices of $\bm \varPsi_1$, $\bm \varPi_1$ in (\ref{U1}), and $\bm \varPsi_2$, $\bm \varPi_2$ in (\ref{U2}).
Due to the fact that the eigendecomposition of a matrix is unique if and only if all its eigenvalues are different, $\bm \varPsi_1$ and $\bm \varPsi_2$ may not be unique.
In addition, if $T_0 < T$, the matrices $\bm \varPi_1$ and $\bm \varPi_2$ can be chosen in arbitrarily many ways, yielding many non-equivalent SDLC schemes.
In the following lemma, we give the condition under which the SDLC scheme is unique.
\begin{lemma}\label{lemma_U_nonunique}
	If both ${\bm {\hat H}}^H {\bm {\hat H}} + {\bm {\hat G}}^H {\bm {\hat G}}$ and $\bm U_1^H {\bm {\hat H}}^H {\bm {\hat H}} \bm U_1$ are full-rank and have distinct positive eigenvalues, the matrix $\bm U$ generated from (\ref{U1U2}) is unique.
	The proposed SDLC scheme is then unique.
	Otherwise, we can obtain as many $\bm U$'s as we want, each corresponding to a different SDLC scheme.
\end{lemma}
\itshape \textbf{Proof:}  \upshape
See Appendix \ref{prove_lemma_U_nonunique}.
\hfill $\Box$	

\begin{remark}\label{two_SD_equal}
	As shown in (\ref{UHHGGU}), we start the simultaneous diagonalization procedure from ${\bm {\hat H}}^H {\bm {\hat H}} + {\bm {\hat G}}^H {\bm {\hat G}}$ and ${\bm {\hat H}}^H {\bm {\hat H}}$.
	Instead, we can also start from ${\bm {\hat H}}^H {\bm {\hat H}} + {\bm {\hat G}}^H {\bm {\hat G}}$ and ${\bm {\hat G}}^H {\bm {\hat G}}$, and solve (\ref{general_problem_1}) by following similar steps. 
	It can be easily proven by symmetry that the two strategies are equivalent.
\end{remark}

\begin{remark}\label{not_optimal_unique_GSVD}
	Analogous to the proposed SDLC scheme, though declared to be optimal, the GSVD-based algorithm provided in \cite{fakoorian2012optimal} can only get the optimal solution of \cite[(10)]{fakoorian2012optimal} rather than that of the original problem \cite[(4)]{fakoorian2012optimal} (similar to (\ref{optimal_A}) being the optimal solution of (\ref{SD_problem3}) rather than (\ref{general_problem_1})).
	In addition, based on the definition of GSVD (see \cite[Definition~$1$]{khisti2010secure}), it can be easily proven that if both ${\bm {\hat H}}^H {\bm {\hat H}} + {\bm {\hat G}}^H {\bm {\hat G}}$ and $\bm U_1^H {\bm {\hat H}}^H {\bm {\hat H}} \bm U_1$ are full-rank and have distinct positive eigenvalues, i.e., they satisfy the uniqueness condition stated in Lemma~\ref{lemma_U_nonunique}, the proposed SDLC scheme is equivalent to the GSVD-based scheme in terms of the secrecy rate.
	However, as shown by the simulation results, the SDLC scheme involves a much lower computational complexity since it avoids computing GSVD.
\end{remark}

\section{General Gaussian MIMO MAC-WT Channel}
\label{K_UE}

Now we consider the general Gaussian MIMO MAC-WT Channel with multiple users.
The channel model and formulated problem have been provided in Section~\ref{GV_MAC-WT}.
Here we solve problem (\ref{C1_max}) by iteratively applying the SDLC scheme proposed in the previous section.
For convenience, denote
\begin{align}\label{omega12}
\bm \varOmega_{1,k} & = \sum_{i \in {\cal K} \setminus k} \bm H_i \bm F_i \bm H_i^H + \sigma_B^2 \bm I_B, \nonumber\\
\bm \varOmega_{2,k} & = \sum_{i \in {\cal K} \setminus k} \bm G_i \bm F_i \bm G_i^H + \sigma_E^2 \bm I_E.
\end{align}
$R (\bm F_{\cal K})$ in (\ref{R_s_joint_GV}) can be rewritten as
\begin{align}\label{R_s_joint_GV_k}
& R (\bm F_{\cal K}) = \left[\log \left| \bm H_k \bm F_k \bm H_k^H \bm \varOmega_{1,k}^{-1} \!+\! \bm I_B \right| \!+\! \log \left| \bm \varOmega_{1,k} \right| \!-\! B \log \sigma_B^2 \right.\nonumber\\
& \left. -\! \log \left| \bm G_k \bm F_k \bm G_k^H \bm \varOmega_{2,k}^{-1} \!+\! \bm I_E \right| \!-\! \log \left| \bm \varOmega_{2,k} \right| \!+\! E \log \sigma_E^2 \right]^+.
\end{align}
Then, if $\bm F_i, \forall i \in {\cal K} \setminus k$ are fixed, problem (\ref{C1_max}) becomes
\begin{subequations}\label{C1_max_2}
	\begin{align}
	\mathop {\max }\limits_{\bm F_k} \; & \log \left| \bm H_k \bm F_k \bm H_k^H \bm \varOmega_{1,k}^{-1} \!+\! \bm I_B \right| \!-\! \log \left| \bm G_k \bm F_k \bm G_k^H \bm \varOmega_{2,k}^{-1} \!+\! \bm I_E \right| \label{C1_max_2a}\\
	\text{s.t.} \;\; &  {\text {tr}}(\bm F_k) \leq P_k, \label{C1_max_2b}\\
	& \bm F_k \succeq \bm 0, \label{C1_max_2c}
	\end{align}
\end{subequations}
where we omit the $\left[ \cdot \right]^+$ operation in the objective function for convenience.
It is obvious from (\ref{omega12}) that $\bm \varOmega_{1,k} \succ \bm 0$ and $\bm \varOmega_{2,k} \succ \bm 0$.
Hence, (\ref{C1_max_2}) can be solved by employing the SDLC scheme proposed in the previous section.
Problem (\ref{C1_max}) can then be solved by iteratively considering (\ref{C1_max_2}) for different users.
The detailed steps are summarized in Algorithm~\ref{SDLC}.
Note that as stated in Remark~\ref{not_optimal}, the SDLC scheme does not necessarily output the optimal solution of (\ref{C1_max_2}).
To guarantee the convergence of Algorithm~\ref{SDLC}, we calculate the new $R (\bm F_{\cal K})$ in each iteration and update ${\bm F}_k$ only if $R (\bm F_{\cal K})$ increases.
\begin{algorithm}[h]
	\begin{algorithmic}[1]
		\caption{SDLC algorithm for solving problem (\ref{C1_max})}
		\State Initialize ${\bm F}_{\cal K}$.
		\Repeat
		\For{$k = 1:K$}
		\State Solve problem (\ref{C1_max_2}) by the SD-based scheme.
		\State Calculate $R (\bm F_{\cal K})$ and update ${\bm F}_k$ if $R (\bm F_{\cal K})$ increases.
		\EndFor
		\Until ${\bm F}_{\cal K}$ converges
		\label{SDLC}
	\end{algorithmic}
\end{algorithm}

We now analyze the complexity of Algorithm~\ref{SDLC}.
For conciseness, we assume equal number of antennas for all users, i.e., $T_k = T, \forall k \in {\cal K}$. 
People can also use $\max \{ T_k, \forall k \in {\cal K} \}$ instead to evaluate the complexity.
In each iteration, as shown in the previous section, the optimization of $\bm F_k$ involves matrix multiplications and eigendecompositions, which yield a complexity of ${\cal O}\left( T^3 \right)$.
In addition, the bisection search used in (\ref{optimal_A}) requires a complexity of ${\cal O}\left( T \log \left( \frac{1}{\epsilon} \right) \right)$, where $\epsilon$ is the convergence tolerance of the bisection searching method.
Let $L$ denote the number of outer iterations of Algorithm~\ref{SDLC}.
Then, the overall complexity of of Algorithm~\ref{SDLC} is ${\cal O}\left( L\left( K\left( T^3 + T \log \left( \frac{1}{\epsilon} \right) \right) \right) \right)$.


\section{Simulation Results}
\label{simulation}

\begin{figure}
	\centering
	\includegraphics[scale=0.50]{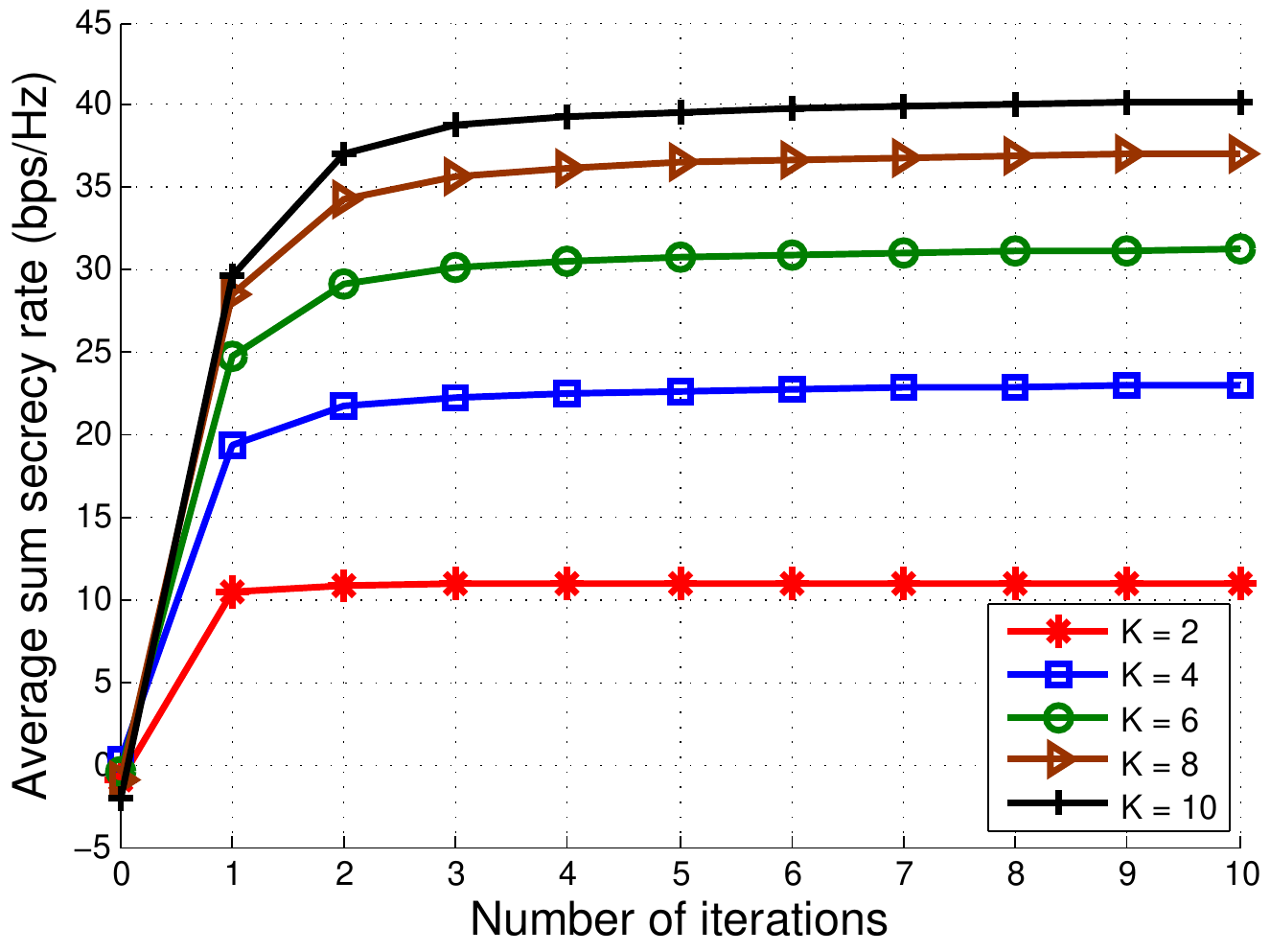}
	\vspace{-1em}
	\caption{Convergence behaviors of the SDLC scheme with $T=4$, $B=E=8$, and $P=10$ dBm.}
	\label{Iteration}
	\vspace{-1.1em}
\end{figure}

\begin{figure}
	\centering
	\includegraphics[scale=0.50]{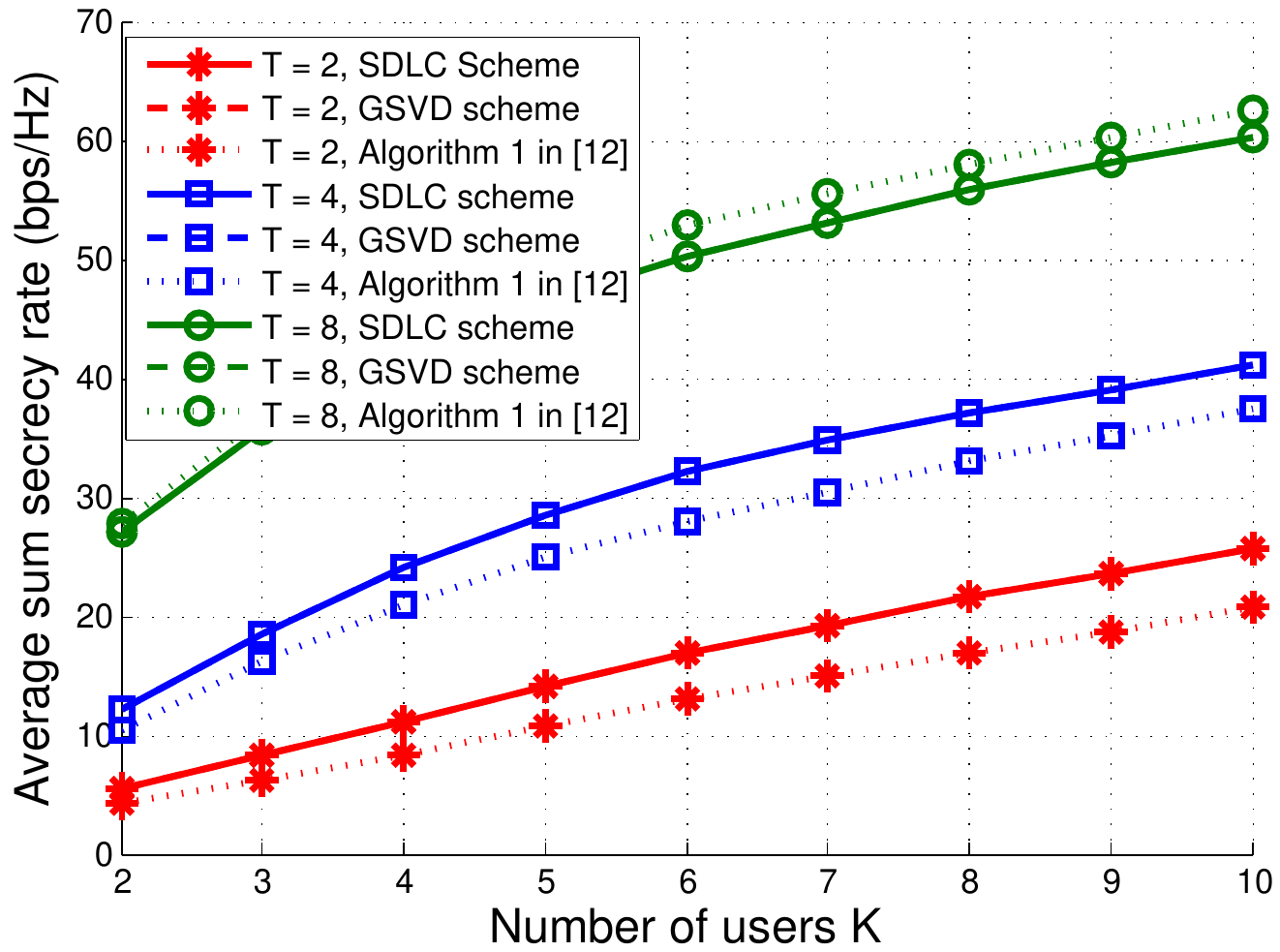}
	\vspace{-1em}
	\caption{Average sum secrecy rate obtained by different schemes with $B=E=8$ and $P=10$ dBm.}
	\label{SSR_VS_TK}
	\vspace{-1.1em}
\end{figure}
In this section, simulation results are presented to evaluate the performance of the proposed algorithms.
We consider an isolated circular-cell network with a radius of $500$ meters.
The base station or Bob is located at the center and an Eve is evenly distributed in the cell.
All mobile users are distributed uniformly in the cell and it is assumed that no user
is closer to Bob than $20$ meters. 
For convenience, equal maximum power constraint, number of antennas at all users, and noise power at Bob and Eve, are assumed, i.e., $P_k = P$, $T_k = T, ~  \forall k\in {\cal K}$, and $\sigma_B^2 = \sigma_E^2 = \sigma^2$. 
The pathloss exponent and the standard deviation of log-normal shadowing fading are respectively set to be $3.7$ and $8$ dB \cite{access2010further}. 
The noise power is $\sigma^2 = -100$ dBm.
All simulation results are obtained by averaging over $1000$ independent channel realizations, and each channel realization is obtained by generating a random
user distribution as well as a random set of fading coefficients.

Fig.~\ref{Iteration} illustrates the convergence behaviors of the proposed SDLC scheme.
It can be seen that the average sum secrecy rate increases greatly during the iterative process and converges rapidly for different configurations of $K$, which shows the significant advantages of the scheme.
When implementing the SDLC scheme in the following, we perform $10$ outer iterations.

\begin{figure}
	\centering
	\includegraphics[scale=0.50]{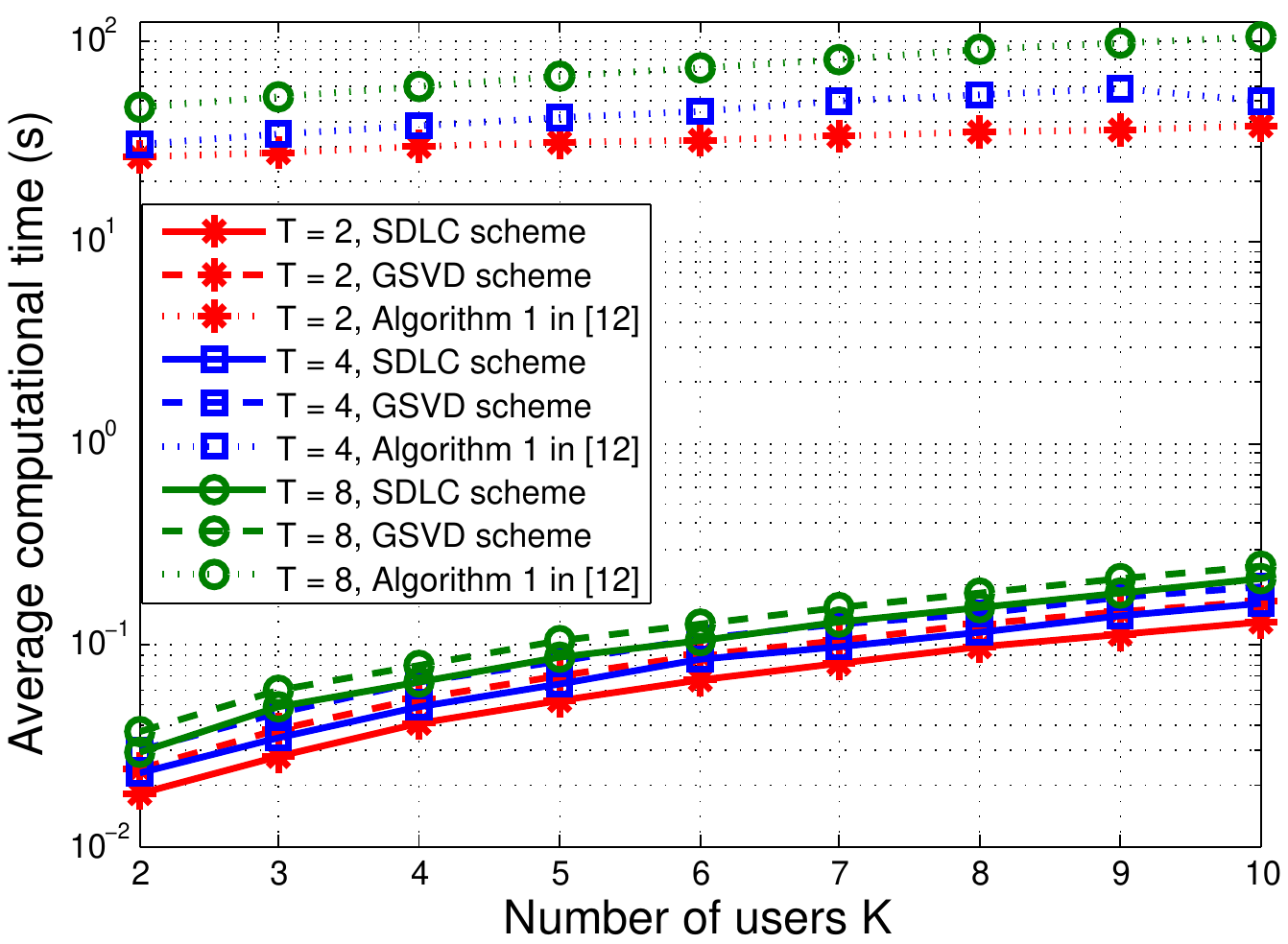}
	\vspace{-1em}
	\caption{Average computational time of different schemes for each channel realization with $B=E=8$ and $P=10$ dBm.}
	\label{time_VS_TK}
	\vspace{-1.1em}
\end{figure}

In Fig.~\ref{SSR_VS_TK} and Fig.~\ref{time_VS_TK}, we compare the SDLC scheme with \cite[Algorithm~$1$]{xu2022achievable} in terms of the secrecy performance and computational complexity.
The results obtained by iteratively applying the GSVD scheme proposed in \cite{fakoorian2012optimal} to solve (\ref{C1_max}) are also depicted as a metric.
We start from the same random initial point when executing different methods and run $10$ outer iterations for the GSVD scheme.
As shown by \cite[Fig.~$2$]{xu2022achievable}, using \cite[Algorithm~$1$]{xu2022achievable}, the average sum secrecy rate increases greatly at the beginning, but then converges quite slowly.
Considering its high complexity, we perform $20$ outer iterations when implementing \cite[Algorithm~$1$]{xu2022achievable}.
In these two figures we vary $K$ and $T$ since they have a significant influence on the computational complexity.

\begin{figure}
	\centering
	\includegraphics[scale=0.50]{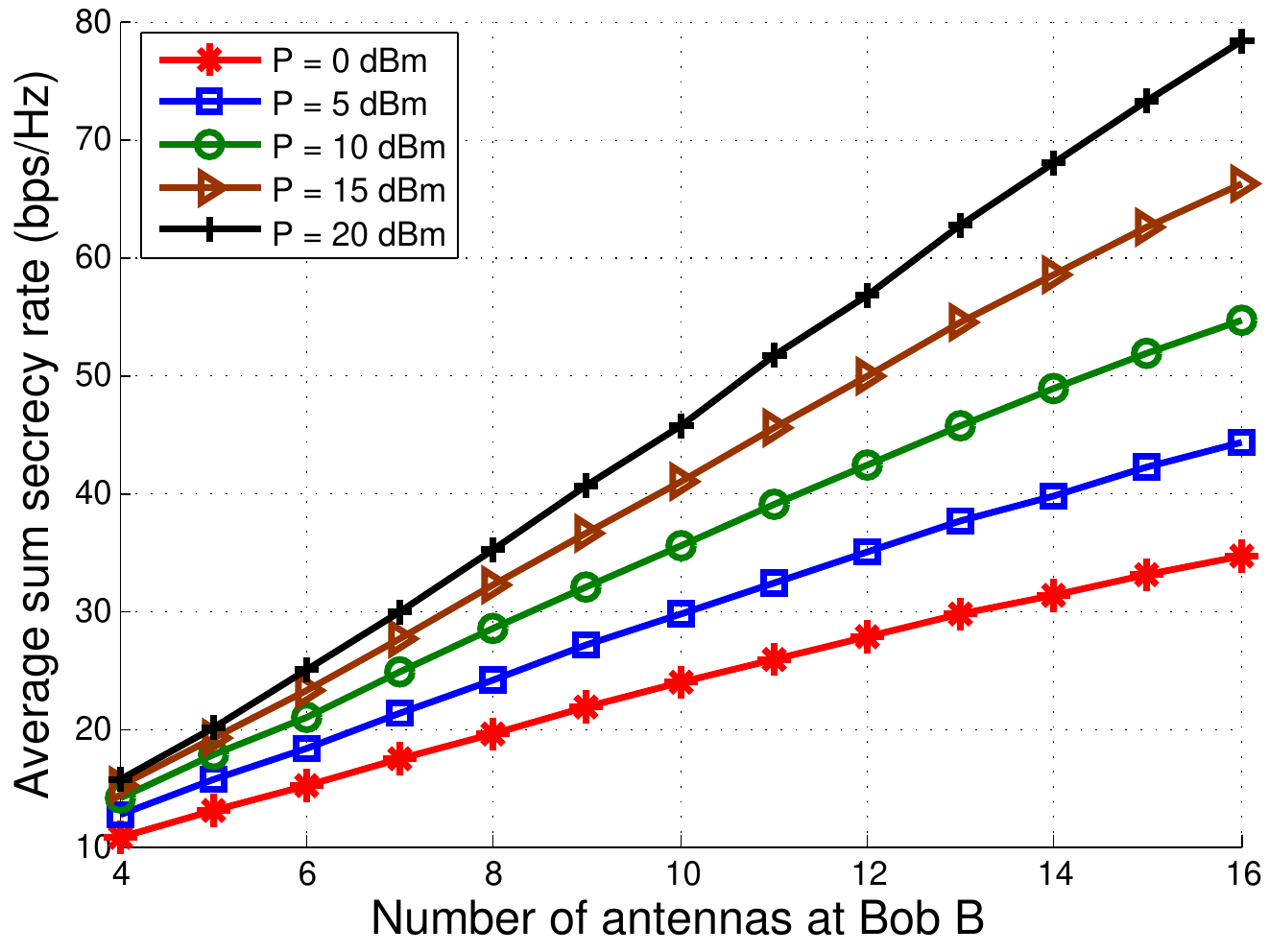}
	\vspace{-1em}
	\caption{Average sum secrecy rate versus the number of antennas at Bob with $K=5$, $T=4$, and $E=8$.}
	\label{SSR_VS_B}
	\vspace{-1.1em}
\end{figure}

As expected, it can be seen from Fig.~\ref{SSR_VS_TK} and Fig.~\ref{time_VS_TK} that both the sum secrecy rate and time cost increase with $K$ and $T$.
Fig.~\ref{SSR_VS_TK} shows that the SDLC scheme has a similar secrecy performance in contrast to \cite[Algorithm~$1$]{xu2022achievable} and the curves obtained by the SDLC and GSVD methods coincide completely, i.e., they have exactly the same secrecy performance.
This is because here both $B$ and $E$ are no smaller than $T$.
${\hat {\bm H}}^H {\hat {\bm H}} + {\hat {\bm G}}^H {\hat {\bm G}}$ and $\bm U_1^H {\hat {\bm H}}^H {\hat {\bm H}} \bm U_1$ then in general are full-rank and have distinct positive eigenvalues.
As explained in Remark~\ref{not_optimal_unique_GSVD}, the proposed SDLC scheme in this case is equivalent to the GSVD method in terms of the secrecy rate.
Nevertheless, as shown by Fig.~\ref{time_VS_TK}, using eigendecomposition instead of the GSVD decomposition, the proposed SDLC scheme reduces the computational time by at least $20 \%$ compared with the GSVD method.
As for \cite[Algorithm~$1$]{xu2022achievable}, it requires over $300$ times of the runtime in contrast to the SDLC scheme, making it impractical to implement this method when $K$ or $T$ is large.

\begin{figure}
	\centering
	\includegraphics[scale=0.50]{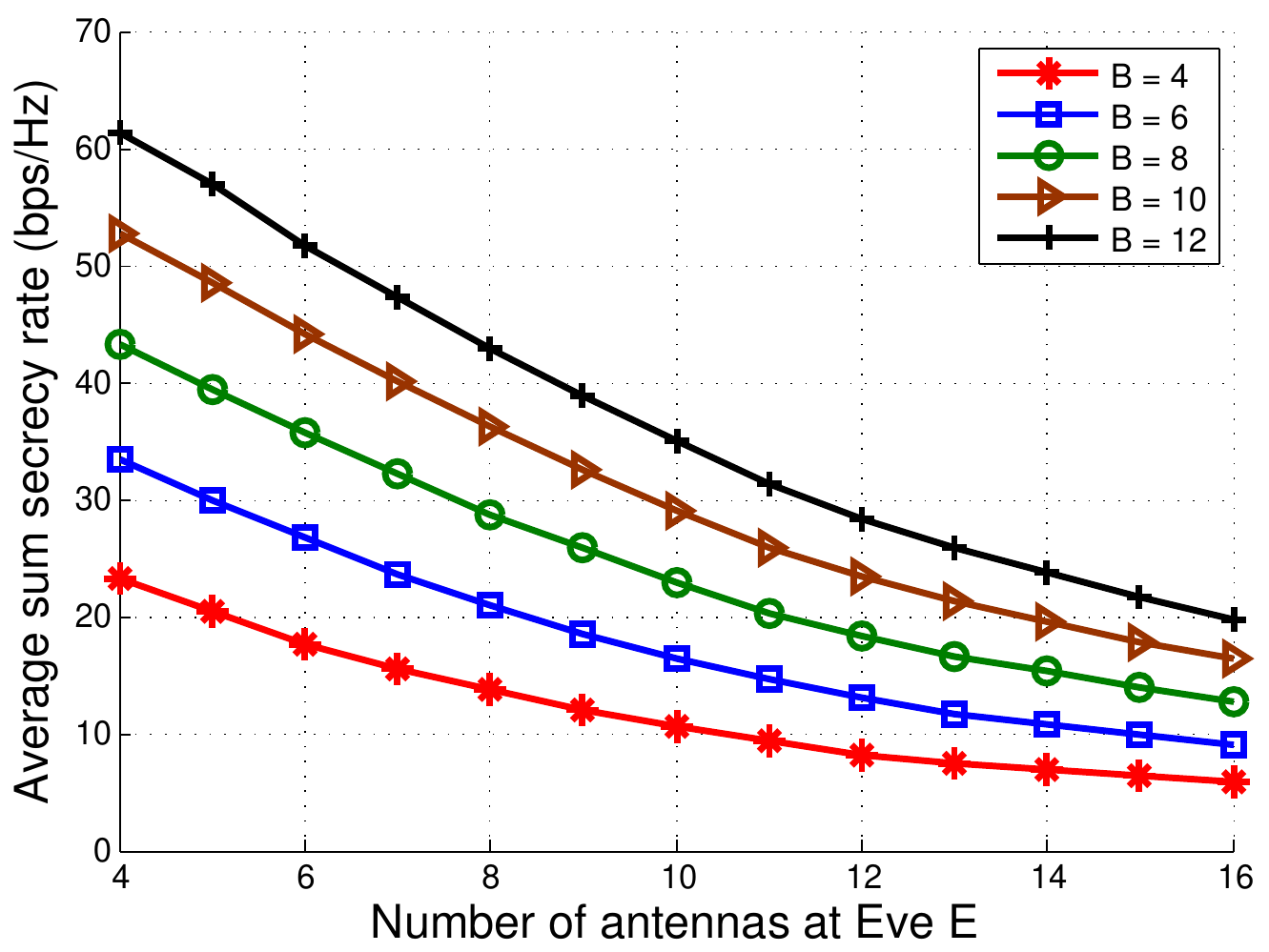}
	\vspace{-1em}
	\caption{Average sum secrecy rate versus the number of antennas at Eve with $K=5$, $T=4$, and $P=10$ dBm.}
	\vspace{-1.1em}
	\label{SSR_VS_E}
\end{figure}
In Fig.\ref{SSR_VS_B} and Fig.~\ref{SSR_VS_E}, we investigate the effect of parameters $B$, $P$, and $E$. 
As expected, the average sum secrecy rate increases with $B$ as well as $P$, and reduces with $E$.


\section{Conclusions}
\label{section6}

This paper has studied the sum secrecy rate maximization problem for a Gaussian MIMO MAC-WT channel.
Due to its high complexity, we did not want to use the conventional MM-based scheme.
Hence, we first proposed a SDLC scheme to maximize the secrecy rate of a single-user wiretap channel and then iteratively optimized the covariance matrices of all users.
Simulation results have confirmed the efficiency and shown that in contrast to the MM-based and GSVD approaches, the proposed SDLC scheme achieves similar secrecy performance but requires much lower computational complexity.

\section*{Acknowledgments}
This work was supported by the European Union's Horizon 2020 Research and Innovation Programme under Marie Skłodowska-Curie Grant No. 101024636 and the Alexander von Humboldt Foundation.

\appendices

\section{Proof of Theorem~\ref{theorem_optimal_a}}
\label{prove_theorem_optimal_a}
As shown in (\ref{SDHHGG}), the last $T - T_0$ diagonal entries of $\bm U^H {\bm {\hat H}}^H {\bm {\hat H}} \bm U$ and $\bm U^H {\bm {\hat G}}^H {\bm {\hat G}} \bm U$ are $0$.
Hence, in the optimal case, we have
\begin{equation}\label{lT0T}
a_t^* = 0, ~\forall~ T_0 + 1 \leq t \leq T.
\end{equation}
Denote ${\cal T}_0 = \{1, \cdots, T_0\}$ and the objective function (\ref{SD_problem3_a}) by
\begin{equation}
r (a_{{\cal T}_0}) = \sum_{t=1}^{T_0} \left[ - \ln (\rho_t a_t + 1) + \ln ((1 - \rho_t) a_t + 1) \right].
\end{equation}
Its first-order partial derivation over $a_t$ is 
\begin{equation}\label{first-deri}
\frac{\partial r}{\partial a_t} = \frac{1 - 2 \rho_t}{(\rho_t a_t + 1)\left[ (1 - \rho_t) a_t + 1 \right]},
\end{equation}
which shows that if $0 \leq \rho_t \leq \frac{1}{2}$, $r (a_{{\cal T}_0})$ is non-decreasing with respect to (w.r.t.) $a_t$.
Since $a_t$ is non-negative, its optimal value is thus
\begin{equation}\label{lt2}
a_t^* = 0, ~{\text {if}}~0 \leq \rho_t \leq \frac{1}{2}.
\end{equation}
Denote set ${\cal J} = \{ t| 1 \leq t \leq T_0,~ \frac{1}{2} < \rho_t \leq 1 \}$.
Then, in the optimal case, $r (a_{{\cal T}_0})$ can be simplified as
\begin{equation}\label{r}
r (a_{\cal J}) = \sum_{t \in {\cal J}} \left[ - \ln (\rho_t a_t + 1) + \ln ((1 - \rho_t) a_t + 1) \right],
\end{equation}
and problem (\ref{SD_problem3}) becomes
\begin{subequations}\label{SD_problem4}
	\begin{align}
	\mathop {\min }\limits_{a_{\cal J}} \quad & r (a_{\cal J}) \label{SD_problem4_a}\\
	\text{s.t.} \quad\; &  a_t \geq 0, ~\forall~ t \in {\cal J}, \label{SD_problem4_b}\\
	& \sum_{t \in {\cal J}} \left\| \bm u_t \right\|^2 a_t \leq P. \label{SD_problem4_c}
	\end{align}
\end{subequations}
Since $\frac{1}{2} < \rho_t \leq 1, \forall~ t \in {\cal J}$, the second-order partial derivation of $r (a_{\cal J})$ over $a_t$ satisfies 
\begin{align}\label{second_deri}
\frac{\partial^2 r}{\partial a_t^2} & = \frac{ (2 \rho_t - 1) \left[ (1 - \rho_t) (2 \rho_t a_t + 1) + \rho_t \right] }{(\rho_t a_t + 1)^2 \left[ (1 - \rho_t) a_t + 1 \right]^2} \nonumber\\
& > 0, ~\forall~ t \in {\cal J}.
\end{align}
(\ref{SD_problem4}) is thus a convex problem.
Due to the affine constraints, the strong duality holds for this problem and its optimal solution could be obtained by checking the KKT condition of its dual problem.
Attaching a Lagrange multiplier $\beta$ to the constraint (\ref{SD_problem4_c}), we get the following Lagrange function
\begin{align}\label{Lagrange_2}
{\cal L} (a_{\cal J}, \beta) & \!=\! \sum_{t \in {\cal J}}\! \left[ -\! \ln (\rho_t a_t \!+\! 1) \!+\! \ln ((1 \!-\! \rho_t) a_t \!+\! 1) \!+\! \beta \left\| \bm u_t \right\|^2 \!a_t \right]\nonumber\\
& - \beta P.
\end{align}
By checking the first-order optimality condition, we know that for any $t \in {\cal J}$,
\begin{equation}\label{lt3}
a_t^* \!=\! \left\{\!\!\!
\begin{array}{ll}
\left[ \frac{1}{\beta^* \left\| \bm u_t \right\|^2} -1 \right]^+, ~{\text {if}}~ \rho_t = 1, \vspace{0.3em} \\
\frac{\left[ - 1 + \sqrt{1 - 4 \rho_t (1 - \rho_t) \left( 1 + \frac{1 - 2 \rho_t}{\beta^* \left\| \bm u_t \right\|^2} \right)} \right]^+}{2 \rho_t (1 - \rho_t) }, {\text {if}}~ \frac{1}{2} \!<\! \rho_t \!<\! 1.
\end{array} \right.
\end{equation}
It can be easily verified that $a_t$ in (\ref{lt3}) monotonically decreases with $\beta$.
Hence, the optimal $\beta^*$ can be found using the bisection searching method such that the constraint (\ref{SD_problem4_c}) holds with equality.
Combining (\ref{lT0T}), (\ref{lt2}), and (\ref{lt3}), we get (\ref{optimal_A}).
This completes the proof.

\section{Proof of Lemma~\ref{lemma_U_nonunique}}
\label{prove_lemma_U_nonunique}
As is well known, if a matrix has distinct eigenvalues, its eigendecomposition is unique (under the convention that if the eigenvalues are sorted in descending order). 
Otherwise, if any two or more eigenvectors share the same eigenvalue, then any set of orthogonal vectors lying in their span are also eigenvectors with that eigenvalue, and we could equivalently choose a unitary matrix using those eigenvectors. 
Therefore, if both ${\bm {\hat H}}^H {\bm {\hat H}} + {\bm {\hat G}}^H {\bm {\hat G}}$ and $\bm U_1^H {\bm {\hat H}}^H {\bm {\hat H}} \bm U_1$ are full-rank and have distinct eigenvalues, $\bm \varPsi_1$ and $\bm \varPsi_2$ are unique and there is no need to add $\bm \varPi_1$ and $\bm \varPi_2$.
$\bm U_1$, $\bm U_2$, and $\bm U$, which are respectively generated from (\ref{U1}), (\ref{U2}), and (\ref{U1U2}), are thus unique.
The proposed SDLC scheme is then unique.

If ${\bm {\hat H}}^H {\bm {\hat H}} + {\bm {\hat G}}^H {\bm {\hat G}}$ or $\bm U_1^H {\bm {\hat H}}^H {\bm {\hat H}} \bm U_1$ is full-rank but has two or more identical eigenvalues, as stated above, we can get many choices of $\bm \varPsi_1$ or $\bm \varPsi_2$.
On the other hand, if ${\bm {\hat H}}^H {\bm {\hat H}} + {\bm {\hat G}}^H {\bm {\hat G}}$ or $\bm U_1^H {\bm {\hat H}}^H {\bm {\hat H}} \bm U_1$ is a defective matrix, since $\bm \varPi_1$ and $\bm \varPi_2$ can be any square matrix of dimension $T - T_0$, we could construct as many $\bm U_1$ and $\bm U_2$ as we want from (\ref{U1}) and (\ref{U2}).
Note that the SDLC scheme is determined by $\bm U$.
In these cases, we can get many different SDLC schemes.
Lemma~\ref{lemma_U_nonunique} is then proven.

\bibliographystyle{IEEEtran}
\bibliography{IEEEabrv,Ref}

\end{document}